# HERITRACE: A User-Friendly Semantic Data Editor with Change Tracking and Provenance Management for Cultural Heritage Institutions

*Arcangelo Massari[1] [1] and Silvio Peroni[2] [1]*

[1] Research Centre for Open Scholarly Metadata, Department of Classical Philology and Italian Studies, University of Bologna, Bologna, Italy

## Abstract (English)

HERITRACE is a data editor designed for galleries, libraries, archives and museums, aimed at simplifying data curation while enabling non-technical domain experts to manage data intuitively without losing its semantic integrity. While the semantic nature of RDF can pose a barrier to data curation due to its complexity, HERITRACE conceals this intricacy while preserving the advantages of semantic representation. The system natively supports provenance management and change tracking, ensuring transparency and accountability throughout the curation process. Although HERITRACE functions effectively out of the box, it offers a straightforward customization interface for technical staff, enabling adaptation to the specific data model required by a given collection. Current applications include the ParaText project, and its adoption is already planned for OpenCitations. Future developments will focus on integrating the RDF Mapping Language (RML) to enhance compatibility with non-RDF data formats, further expanding its applicability in digital heritage management.

## Abstract (Italian)

HERITRACE è un editor di dati semantici progettato per gallerie, biblioteche, archivi e musei, con l'obiettivo di semplificare la curatela dei dati consentendo agli esperti non tecnici di gestirli in modo intuitivo senza compromettere la loro integrità semantica. Sebbene le tecnologie del Web Semantico possano rappresentare una barriera alla curatela dei dati a causa della loro complessità, HERITRACE offusca questa complessità preservando al contempo i vantaggi della rappresentazione semantica. Il sistema supporta nativamente la gestione della provenance e il tracciamento delle modifiche, garantendo trasparenza e responsabilità durante il processo di curatela. Pur funzionando efficacemente fin dal primo utilizzo, HERITRACE offre un'interfaccia di personalizzazione semplice per il personale tecnico, consentendo l'adattamento al modello di dati specifico richiesto da una determinata collezione. Tra le applicazioni attuali figura il progetto ParaText, e l'adozione è già prevista per OpenCitations. Gli sviluppi futuri si concentreranno sull'integrazione del RDF Mapping Language (RML) per migliorare la compatibilità con i formati di dati non RDF, ampliando ulteriormente la sua applicabilità nella gestione del patrimonio digitale.

---

[1] arcangelo.massari@unibo.it
[2] silvio.peroni@unibo.it

# 1 Introduction

Traditionally, GLAM (galleries, libraries, archives, and museums) experts have relied on their interpretative skills and domain knowledge to curate metadata. However, the digitization of cultural heritage data has introduced new challenges, including the representation of data in various machine-readable and semantic-actionable formats and their preservation in heterogeneous databases and repositories. This scenario has created significant barriers for domain experts lacking technical expertise, particularly in Semantic Web technologies.

The widespread adoption of Semantic Web technologies in the GLAM sector underscores their growing relevance. Numerous prominent institutions, including the Library of Congress (Laurence 2013), Bibliothèque nationale de France (da Silveira et al. 2020), and Deutsche National Bibliothek (Hannemann and Kett 2010) have embraced the Linked Open Data (LOD) paradigm to enhance the findability, accessibility, interoperability and reusability of their collections (Wilkinson et al. 2016). RDF (Pan 2009) has been instrumental in these efforts, enabling seamless integration of diverse data formats and fostering computational analysis. Candela highlights how RDF facilitates not only data structuring but also its reuse and enrichment, making it an essential middleware technology for GLAM collections (Candela 2023). Additional studies emphasize the pivotal role of RDF and LOD in creating interoperable datasets and fostering global collaboration across institutions (Alexiev 2018; Candela et al. 2022; Ranjgar et al. 2024).

While Semantic Web technologies have unlocked new possibilities, they have also led to a paradoxical situation. On the one hand, these technologies have made human intervention more critical due to the semantic interpretation of data that cannot be completely automated. On the other hand, they have limited the number of curators to those who are experts in the Semantic Web, thereby creating challenges in workforce scalability and accessibility. This duality has resulted in contrasting scenarios within the GLAM domain. Some collections have embraced Semantic Web technologies, requiring staff with advanced technical expertise for maintenance, while others have avoided these technologies to prevent curatorial complexities.

Two examples illustrate these divergent paths: the FICLIT Digital Library (ADLab - Laboratorio Analogico Digitale and /DH.arc - Digital Humanities Advanced Research Centre 2022) and OpenCitations (Peroni and Shotton 2020), both managed by the University of Bologna. The FICLIT Digital Library, relying on Omeka S, faces limitations due to its simplistic semantic tools and lack of SPARQL query capabilities, leading to challenges in change tracking and transparent provenance management. In contrast, OpenCitations fully embraces Semantic Web technologies but grapples with the issue of incorrect or missing data, a problem that requires human discernment for correction.

The central problem that arises from these scenarios is the gap between the complex digital technologies and the domain expertise of GLAM professionals. This gap hinders effective data curation and limits the potential of digital collections to represent and disseminate cultural heritage accurately and comprehensively. The solution proposed in this project is the development of a framework that facilitates domain experts without skills in Semantic Web technologies in enriching and editing such semantic data intuitively, irrespective of the underlying ontology model and the technologies adopted for storing such data.

The challenges are manifold. A critical goal is to create a system that is user-friendly for several kinds of end-users, including librarians, museologists, gallery curators, archivists, administrators and IT professionals who are tasked with setting up and maintaining the framework. Another significant challenge is provenance management. In the context of GLAM institutions, where the historical and source context of data is paramount, a data management system must accurately track and document the responsible agents and primary data sources. Change-tracking is also a fundamental requirement. The system needs to efficiently monitor and record all modifications to the data, allowing for transparency and accountability in the curation process. Customization is a further challenge that a data management system for cultural heritage must address. Recognizing that different GLAM domains have unique requirements for how resources are represented and managed, a customizable interface should be tailored to various data models, enabling the representation of diverse resource types according to specific domain needs. Finally, interfacing with pre-existing data presents a substantial challenge, as GLAM institutions often already possess vast collections, which organize their data through the adoption of different data models. This requirement is particularly important for ensuring that the transition to a new data management system is smooth and does not disrupt the ongoing operations of the institution.

To address all these problems, we have developed a novel Web-based tool called HERITRACE (Heritage Enhanced Repository Interface for Tracing, Research, Archival Curation, and Engagement). HERITRACE has been designed with five primary objectives: (1) providing a user-friendly interface for domain experts to interact with semantic data without technical knowledge; (2) implementing comprehensive provenance management to document metadata modifications; (3) delivering robust change-tracking capabilities for reconstruction of previous data states; (4) offering flexible customization through standardized languages rather than proprietary solutions; and (5) facilitating seamless integration with pre-existing RDF data collections. Through these objectives, HERITRACE aims to bridge the gap between sophisticated semantic technologies and the practical needs of cultural heritage professionals.

This article extends a previous publication – included in the Book of Abstracts of AIUCD 2024 (Massari and Peroni 2024) – adding additional information about the technologies used for implementing HERITRACE, the comparison with existing tools, and the features and functions it implements.

The rest of this paper is structured as follows: Section 2 examines related work through two perspectives: the adoption of Semantic Web technologies in cultural heritage institutions (Section 2.1) and a critical comparative analysis of existing semantic editors (Section 2.2). Section 3 presents HERITRACE's architecture and key functionalities, particularly its user interface design, provenance management system, and change-tracking mechanisms. Finally, Section 4 evaluates HERITRACE's contributions to the field, discussing how its integrated approach resolves limitations identified in current systems, while outlining future development paths.

## 2  Related Work

Semantic Web technologies are changing how museums, libraries, archives, and galleries manage their collections. Cultural institutions worldwide are increasingly structuring their data as Linked Open Data LOD, making cultural heritage not just more accessible but also more interconnected. This section explores two aspects of this evolution: first, how cultural institutions are embracing these technologies in their digital projects; and second, what tools are available to help domain experts, who may not be technical specialists, easily manage this semantic data.

### 2.1  Adoption of Semantic Web Technologies in GLAM Institutions

The past decade has witnessed a significant transformation in how cultural heritage institutions manage and share their collections, with many organizations embracing Semantic Web technologies and LOD principles. Table 1 showcases the breadth of this adoption across Europe and beyond.

The Österreichische Nationalbibliothek (Rachinger 2008) has converted its authority files and bibliographic records to LOD, making over 1.8 million objects accessible via the Web, including medieval manuscripts, early printed books, and the world's largest collection of papyri. The Biblioteca Nacional de España (Wulff 2024) offers RDF representations of its catalog that connect authors, works, and subjects, providing access to Spanish literary heritage. The Biblioteca Virtual Miguel de Cervantes (Candela et al. 2018) has focused on semantically enriching its digital collections of Hispanic literature, creating connections between literary works, their authors, and their historical context.

The Bibliothèque nationale de France (da Silveira et al. 2020) uses LOD to enhance access to French cultural heritage, aggregating information previously scattered across various databases, spanning from medieval manuscripts to contemporary publications. Similar efforts are visible at the Bibliothèque nationale du Luxembourg (Kremer 2021), which has prioritized the semantic representation of its multilingual collections, while the British National Bibliography (Corine et al. 2017) has focused on publishing bibliographic metadata as LOD, enabling complex queries impossible with traditional catalogs.

The Deutsche National Bibliothek (Hannemann and Kett 2010) provide access to German-language cultural resources, while the Library of Congress offers LOD services that not only cover American cultural heritage but also provide reference points for cultural institutions worldwide through their authority data (Laurence 2013). The National Library of Finland (Hyvönen 2020) has integrated its data with the broader Finnish LOD infrastructure, while the National Library of the Netherlands (Beek et al. 2014) has focused on connecting its historical collections through semantic relationships.

In Italy, a notable initiative is ARCo (the Italian Cultural Heritage Knowledge Graph). Developed through collaboration between the Italian Ministry of Culture and the National Research Council, ARCo transforms the Ministry's official General Catalogue into a rich knowledge graph. With data about 820 thousand cultural entities, it represents a significant step toward making Italy's exceptional cultural patrimony digitally accessible and semantically structured (Carriero et al. 2019).

However, it is the Europeana project (Isaac and Haslhofer 2013) that represents the most ambitious implementation in the field, aggregating and semantically enriching cultural heritage data from institutions across Europe.

These implementations, while successful in making cultural heritage data more accessible and interoperable, highlight a critical need: tools that allow domain experts - librarians, archivists, and curators - to manage and enrich semantic data without requiring extensive technical knowledge. The complexity of current LOD management systems often creates a barrier between cultural heritage professionals and their own data, limiting their ability to maintain and enhance these valuable resources. This situation underscores the importance of developing user-friendly semantic editors that can work with various data models while remaining accessible to non-technical users.

| Institution | URL | Reference |
|---|---|---|
| Österreichische Nationalbibliothek | https://labs.onb.ac.at | (Rachinger 2008; Petz 2023) |
| Biblioteca Nacional de España | http://datos.bne.es | (Wulff 2024) |
| Biblioteca Virtual Miguel de Cervantes | https://data.cervantesvirtual.com | (Candela et al. 2018) |
| Bibliothèque nationale de France | https://data.bnf.fr | (da Silveira et al. 2020) |
| Bibliothèque nationale du Luxembourg | https://data.bnl.lu | (Kremer 2021) |
| British National Bibliography | https://bl.natbib-lod.org/ | (Corine et al. 2017) |
| Italian Ministry of Culture (ARCo project) | https://dati.beniculturali.it/arco-rete-ontologie | (Carriero et al. 2019) |
| Europeana | https://pro.europeana.eu | (Isaac and Haslhofer 2013) |
| Deutsche National Bibliothek | https://www.dnb.de | (Hannemann and Kett 2010) |
| Library of Congress | https://id.loc.gov | (Laurence 2013) |
| National Library of Finland | https://data.nationallibrary.fi | (Hyvönen 2020) |
| National Library of the Netherlands | https://data.bibliotheken.nl | (Beek et al. 2014) |

*Table 1 Overview of LOD repositories published by GLAM organizations, adapted from (Candela 2023).*

## 2.2 Comparative Analysis of Semantic Editors in GLAM

Building on the discussion of Semantic Web adoption, this subsection shifts the focus on semantic editors. A comparative evaluation of systems, including OmekaS (Salarelli 2016), Semantic MediaWiki (Krötzsch, Vrandečić, and Völkel 2006), Research Space (Oldman and Tanase 2018), and CLEF (Daquino et al. 2022), reveals varying strengths and limitations across criteria like user-friendliness, provenance management, change tracking, customization, and compatibility with heterogeneous data sources. These evaluation criteria are based on those used to assess the CLEF system in the related paper. This ensures that our assessment criteria are not only relevant but also consistently applied across similar platforms within the digital heritage domain, as summarized in Table 2.

| Name | User friendly (Users) | User friendly (Admin) | Provenance Mgmt. | Change-tracking | Customization | Heterogeneous data sources |
|---|---|---|---|---|---|---|
| OmekaS | ✓ | ✓ | | | ✓ | |
| Semantic MediaWiki | ✓ | ✓ | ✓ | ✓ | ✓ | |
| Research Space | ✓ | | ✓ | | ✓ | ✓ |
| CLEF | ✓ | ✓ | ✓ | ✓ | | |

*Table 2 Comparison of Data Management System Features for the GLAM Sector.*

OmekaS, recognized for its user-friendly interface, primarily serves museums and educational institutions with its intuitive web-publishing platform. However, it exhibits certain limitations in more complex operational aspects. Notably, OmekaS does not inherently track provenance. This limitation can affect the credibility and traceability of the information presented. Additionally, OmekaS lacks inbuilt change-tracking capabilities. Data interfacing in OmekaS presents another challenge. To import pre-existing data in bulk, users must rely on the CSV Import plugin (Berthereau and Corporation for Digital Scholarship 2015). This plugin necessitates restructuring the original data to fit its specific format with mandatory field names, which can be a cumbersome and time-consuming process. This requirement for data formatting reduces the platform's flexibility in handling heterogeneous data sources.

Semantic MediaWiki significantly enhances the popular MediaWiki platform by integrating semantic capabilities. This blend of features balances user-friendliness for non-technical end-users and the more complex needs of technical administrators. One of the key strengths of Semantic MediaWiki is its customization potential, although it requires a degree of familiarity with both the MediaWiki environment and underlying semantic concepts. In terms of data provenance management, Semantic MediaWiki provides robust support. However, its capabilities for change-tracking are not native to the system but are instead supplemented through the use of external plugins. A notable example is the Semantic Watchlist plugin (WikiTeq 2022), which effectively monitors changes within the wiki. These changes are stored in a relational database rather than in RDF format, which, while practical for tracking purposes, may not align seamlessly with the semantic structure of the data. This discrepancy could potentially restrict the depth of change analysis and the ability to contextualize changes within the semantic framework of the data. Addressing the interfacing with heterogeneous data sources, Semantic MediaWiki initially focused solely on importing OWL ontologies. To broaden its RDF support, the RDFIO extension was introduced (Lampa et al. 2017). This extension enables the loading of RDF triples, but it is confined to the N-Triples format and notably lacks support for named graphs. This limitation is significant as it restricts the platform's adaptability in various environments that may require more complex semantic data structures.

Research Space, tailored for the academic and research community, excels in user-friendliness for end-users, offering diverse data visualization options such as graphs and temporal maps. However, it maintains a level of complexity for administrators, demanding a steep learning curve. The platform requires a solid understanding of HTML, handlebars and other ResearchSpace-specific components for creating templates, which may be cumbersome even for those with technical expertise. In terms of data provenance, Research Space automatically associates data with its source, ensuring traceability and credibility. However, it lacks a change-tracking system, which could limit its effectiveness in environments where monitoring data modifications over time is crucial. Regarding data interfacing, Research Space allows uploading RDF data directly, which is advantageous for projects involving such formats. However, after the data is uploaded, an administrator's intervention is required to customize the interface appropriately to display the items correctly. This aspect indicates that while Research Space can interface with heterogeneous data sources, doing so involves a significant level of programming complexity for system administrators.

CLEF is designed to manage complex digital libraries, archives, and research data, particularly in the humanities. It offers an administrator-friendly interface and focuses on user-friendliness for end-users, making it suitable for a wide range of audiences within its domain. CLEF's provenance management is robust, utilizing named graphs. Moreover, it does feature change tracking capabilities, including synchronization with GitHub, but lacks a direct system to restore previous versions. Expanding on the capabilities of CLEF, it is important to note that this system does not allow for extensive customization. Moreover, unlike some of its counterparts, CLEF is not designed to upload and manage pre-existing RDF data as-is. This limitation is significant because the software is structured to add items one by one from scratch directly through the user interface. This approach, while potentially beneficial for building new databases, limits the platform's ability to seamlessly integrate and manage existing large-scale datasets. Furthermore, even though CLEF does not impose a specific data model, it organizes data in a format akin to nanopublications for managing provenance. This structure means that if a pre-existing triple store is connected, the system is not ready to explore the data without a prior reorganization to make it compatible with CLEF's framework.

The comparative analysis highlights how these systems cater to different needs within the GLAM sector. While tools like OmekaS focus on simplicity and accessibility for non-technical users, they lack advanced features like provenance tracking and change management. Semantic MediaWiki and Research Space offer richer functionalities but come with a steeper learning curve, especially for administrators. CLEF stands out for its robust provenance management and synchronization with external platforms, yet its limited ability to handle pre-existing RDF data restricts its flexibility. These insights underline the need for a balanced approach that combines usability with advanced semantic capabilities, paving the way for tools that can bridge these gaps effectively.

# 3   HERITRACE Functionalities

The user interface (UI) of HERITRACE is designed to cater to the needs of domain experts in the GLAM sector, ensuring that they can interact with complex datasets without requiring technical expertise. The UI is web-based, allowing access from various devices and platforms, including both PCs and smartphones, and supports professionals by enabling them to work across different environments. Access to the system is restricted to pre-authorized personnel, with login credentials managed via ORCID (Haak et al. 2012), ensuring secure and streamlined authentication.

This flexibility and security are demonstrated in HERITRACE's deployment within the ParaText project, which serves as a real-world case study for the system's capabilities. Developed by the University of Bologna, ParaText focuses on managing bibliographic metadata for textual resources. While access to ParaText is restricted to authenticated and authorized personnel, all examples and screenshots illustrating HERITRACE's functionalities in this paper are drawn from its deployment in the ParaText project.

While HERITRACE is not primarily engineered as a comprehensive search tool— unlike specialized systems such as OSCAR (Heibi, Peroni, and Shotton 2019) — it provides a basic interface for visualizing the contents of a dataset. The catalog interface in HERITRACE presents a two-panel layout. On the one hand, users can find a "Categories" panel that lists all available types of objects described in the system, with each category displaying the total number of items it contains. For example, Figure 1 shows various publication types such as "Article in Book" (153 items), "Issue" (25 items), "Journal" (27 items), and so forth. On the other hand, users can view the items belonging to the selected category. The interface includes practical features such as a "Sort by" dropdown menu allowing users to order items by different properties, an "Items per page" selector for controlling the number of displayed results, and pagination controls for navigating through the list of items. Finally, each item in the list is displayed as a clickable link.

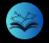

Figure 1 HERITRACE catalog interface.

When a user clicks on an item in the catalog, they are presented with a detailed view of the resource. This interface displays the resource's metadata fields, such as type, identifier, and title as shown in Figure 2, depending on the structure of the specific dataset. Each field is labeled and paired with its current value, allowing users to view or modify the information through inline editing. Intuitive controls, like text inputs and dropdown menus, facilitate straightforward updates, with real-time validation to ensure data consistency. Action buttons at the top of the interface enable users to perform key operations. These include canceling edits without saving changes, deleting the resource, or accessing the "Time Machine" to view and restore previous versions. Users can also expand the metadata by adding new fields, such as additional identifiers or descriptions.

# Pagani, L. (2015). Al crocevia di lingua e letteratura. Il grammatico Filosseno come esegeta di Omero. In: Lemmata. Beiträge zum Gedenken an Christos Theodoridis.

[Cancel Editing] [Delete] [Time Machine]

**Type**

Article in Book ⓘ

expression ⓘ

**Identifier**

doi:10.1515/9783110354348-019

+ Add Identifier

**Title**

Al crocevia di lingua e letteratura. Il grammatico
Filosseno come esegeta di Omero

+ Add Description

*Figure 2 Detailed view of a resource in HERITRACE, showing metadata fields and editing options.*

The "Time Machine" feature in HERITRACE, illustrated in Figure 3, provides users with a timeline for managing the evolution of each entity in the database. Every time an entity is created, modified, or deleted the system generates a new snapshot. These snapshots capture key provenance metadata, including the timestamp of generation, the responsible agent, the primary data source, a description of the entity, and a detailed list of the modifications made.

The timeline interface allows users to view previous versions of a resource, facilitating comparisons and enabling the restoration of earlier states if the current version is no longer suitable. When a user decides to restore an entity to a previous snapshot, HERITRACE automatically updates the resource and ensures that all linked resources are also adjusted accordingly. For example, if a bibliographic resource is restored to a previous version where its associated identifiers or descriptive metadata differ from the current state, HERITRACE ensures that the linked records are reverted to previous snapshots as well. This integrated approach minimizes the risk of inconsistencies and guarantees that the dataset remains coherent and accurate throughout its evolution. The "Time Machine" thus is a critical tool for maintaining transparency, accountability, and reliability in metadata curation and management.

While existing entities are accessible through the main catalog interface, when resources are deleted, they disappear from this catalog but are never permanently removed from the system. This is where Time Vault complements the Time Machine functionality. Time Vault serves as a dedicated "recycle bin" or catalog specifically for deleted entities, making them easily discoverable through a specialized interface. Users can review essential information about deleted resources, such as the deletion timestamp, the responsible agent, and the associated modifications. Additionally, the Time Vault enables quick access to the most recent version of a deleted resource, allowing users to restore it if necessary.

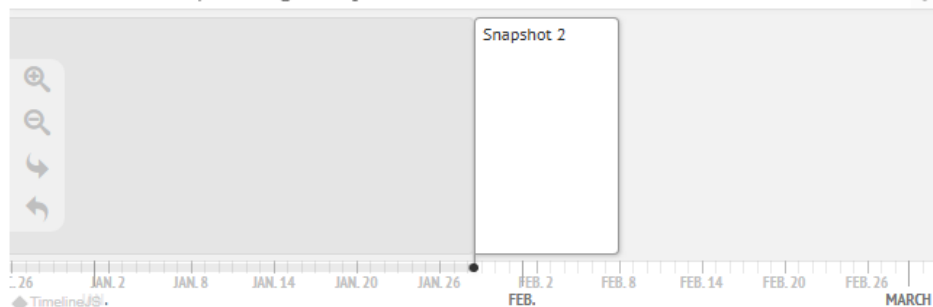

Figure 3 The HERITRACE timeline interface.

In addition to managing existing resources, HERITRACE provides an interface for creating new records, ensuring that curators can enrich their datasets while maintaining the same standards of accuracy and consistency. This process begins with the "New Record" section, where users can easily define new entities and their associated metadata. The system dynamically adjusts the available metadata fields based on the selected type, ensuring compatibility with the dataset's underlying structure. For example, as shown in Figure 4, if "Article" is selected as the entity type, the interface presents fields such as "Identifier," "Title," "Description," "Abstract," "Keyword," "Author," "Editor," "Collaborator," and "Publisher." Users can expand and fill these fields as needed. Each field includes clear labels and helpful tooltips to guide users in entering accurate and relevant metadata.

To ensure the integrity and reliability of the data, HERITRACE incorporates a robust validation system that applies both to the creation of new records and to the modification of existing entities. As users populate or update metadata fields, the system performs real-time checks to validate the entered information against predefined rules and constraints. These constraints, defined through SHACL (Pareti and Konstantinidis 2022), may include conditions such as mandatory fields, value types (e.g., text, numbers, or URIs), or the allowable number of values for a specific property. Validation feedback is provided immediately, highlighting any issues and guiding users to correct errors or omissions before saving changes. This ensures that all new and modified records adhere to the dataset's structural and semantic requirements, minimizing errors and promoting consistency across the collection.

Beyond validation, HERITRACE implements a disambiguation system that activates during entity creation and modification. When users enter metadata such as titles, author names, or identifiers, the system automatically searches for existing entities with similar attributes to prevent unintentional duplication. As shown in Figure 4, when a user types "Franco" in the Given Name field, the system suggests "Franco Montanari [orcid:09110155]" who is already in the database as the author of another journal article. This disambiguation process can be configured through the system's YAML settings, as detailed in Section 3.1. The feature is particularly valuable in large, collaborative collections where multiple curators might otherwise inadvertently create duplicate entries for the same underlying entity.

*Figure 4 HERITRACE interface for creating a new resource. As shown in the figure, when a user types "Franco" in the Given Name field, the system suggests "Franco Montanari" who is already in the database as the author of another journal article, demonstrating the disambiguation system in action*

## 3.1 Configurations and customization

HERITRACE is designed to function seamlessly with any RDF-based data set, allowing institutions to deploy the system without the need for extensive initial configuration. However, for those who wish to tailor the system to their specific needs, HERITRACE offers extensive configuration options to enhance its functionality and user experience. This subsection delves into a more technical aspect of HERITRACE, focusing on the role of the technical staff or configurator rather than the domain expert. The system is designed to simplify the configurator's task by allowing customization through well-known languages, particularly SHACL (Pareti and Konstantinidis 2022) and YAML (Ben-Kiki, Evans, and Ingerson 2009).

SHACL (Shapes Constraint Language) is a W3C standard used to validate RDF data against a set of conditions. It enables the definition of constraints on data models, ensuring that the data adheres to specific structural and semantic rules. In HERITRACE, SHACL is employed to define the data model, specifying classes, properties, and constraints for each entity type. This ensures data consistency and integrity across the dataset. In HERITRACE, SHACL determines not only which properties belong to a particular class but also how these properties should be managed and validated within the system.

For instance, if the SHACL shape related to the class `fabio:JournalArticle` states that the `prism:publicationDate` property can have a data type of `xsd:date`, `xsd:gYearMonth`, or `xsd:gYear`, the system will present a dropdown menu allowing users to select one of these date types. Depending on the selection, the appropriate input method will be displayed: a calendar for a full date, a month-year selector for `xsd:gYearMonth`, and a simple number input for `xsd:gYear`.

As said, SHACL also defines cardinality constraints using `sh:minCount` and `sh:maxCount`, which dictate how many times a property can be added. For example, a property like `dcterms:title` might have a max count of 1, indicating it can only appear once, whereas properties like `datacite:hasIdentifier` might have no upper limit, allowing multiple entries.

Beyond basic property definitions, SHACL supports sophisticated validation mechanisms. Regular expression patterns can be specified to validate specific formats, such as a DOI (Digital Object Identifier). By embedding a pattern within the SHACL definition, the system can ensure that the DOI adheres to its expected format. Additionally, SHACL allows for custom error messages to be specified, providing clear feedback if the input does not match the required pattern.

Complementing SHACL, HERITRACE uses YAML configuration files to control the visual presentation of data. Here are some examples of YAML configurations used to customize the HERITRACE interface. Listing 1 illustrates how administrators can configure the visual representation of a Journal Article entity in the HERITRACE user interface.

```yaml
1.  - class: "http://purl.org/spar/fabio/JournalArticle"
2.    priority: 1
3.    shouldBeDisplayed: true
4.    displayName: "Journal Article"
5.    fetchUriDisplay: |
6.      PREFIX dcterms: <http://purl.org/dc/terms/>
7.      PREFIX fabio: <http://purl.org/spar/fabio/>
8.      PREFIX foaf: <http://xmlns.com/foaf/0.1/>
9.      PREFIX pro: <http://purl.org/spar/pro/>
10.     SELECT ?display
11.     WHERE {
12.       [[uri]] dcterms:title ?title .
13.       OPTIONAL {SELECT (GROUP_CONCAT(?authorName; SEPARATOR = " & ")
14.         AS ?authorList)
15.         WHERE {
16.           [[uri]] pro:isDocumentContextFor ?authorRole .
17.           ?authorRole pro:withRole pro:author ;
18.                   pro:isHeldBy ?author .
19.           ?author foaf:familyName ?authorName .
20.         }
21.       }
22.       BIND(CONCAT(
23.         COALESCE(?authorList, ""),
24.         IF(BOUND(?authorList), ". ", ""),
25.         ?title
26.       ) AS ?display)
27.     }
28.   displayProperties:
29.     - property: "http://www.w3.org/1999/02/22-rdf-syntax-ns#type"
30.       displayName: "Type"
31.       shouldBeDisplayed: true
32.       supportsSearch: false
33.     - property: "http://purl.org/dc/terms/title"
34.       displayName: "Title"
35.       shouldBeDisplayed: true
36.       inputType: "textarea"
37.       supportsSearch: true
38.       minCharsForSearch: 2
39.       searchTarget: "self"
40.     - property: "http://purl.org/spar/datacite/hasIdentifier"
41.       displayName: "Identifier"
42.       shouldBeDisplayed: true
43.       supportsSearch: true
44.       searchTarget: "parent"
45.       fetchValueFromQuery: |
46.         PREFIX datacite: <http://purl.org/spar/datacite/>
47.         PREFIX literal:
48.                 <http://www.essepuntato.it/2010/06/literalreification/>
49.         SELECT (CONCAT(STRAFTER(STR(?scheme), "datacite/"), ":",
50.                 ?literal) AS ?id) ?identifier
51.         WHERE {
52.           [[subject]] datacite:hasIdentifier ?identifier.
53.           ?identifier datacite:usesIdentifierScheme ?scheme;
54.                   literal:hasLiteralValue ?literal.
55.         }
56.     - property: "http://purl.org/vocab/frbr/core#partOf"
57.       displayName: "Issue"
58.       shouldBeDisplayed: true
59.       supportsSearch: true
60.       fetchValueFromQuery: |
61.         PREFIX frbr: <http://purl.org/vocab/frbr/core#>
62.         PREFIX dcterms: <http://purl.org/dc/terms/>
63.         PREFIX fabio: <http://purl.org/spar/fabio/>
64.         SELECT ?containerName ?container
65.         WHERE {
66.           [[subject]] frbr:partOf+ ?container.
67.           ?container a fabio:JournalIssue.
68.           ?container dcterms:title ?containerName.
69.         }
70.
```

*Listing 1 YAML configuration for a Journal Article entity in ParaText.*

The configuration in Listing 1 contains several components for customizing how Journal Articles are displayed and interacted with in HERITRACE:

- The `class` attribute specifies the RDF class (`fabio:JournalArticle`) to which this configuration applies.
- The `priority` attribute determines which configuration to apply when an entity has multiple RDF types. Lower numbers take precedence, so an entity with both `fabio:Expression` and `fabio:JournalArticle` types will use the `JournalArticle` configuration if it has a lower priority number.
- The `shouldBeDisplayed` attribute controls whether this entity type appears in the catalog interface and whether it can be created through the "New Record" interface. If set to false, users cannot create new entities of this type.
- The `displayName` attribute provides a user-friendly label for the class itself, showing "Journal Article" in the interface instead of the RDF class URI `http://purl.org/spar/fabio/JournalArticle`.
- The `fetchUriDisplay` property contains a SPARQL query that transforms each entity's URI into a human-readable label. In this example, it generates a display string combining author names and the article title, which replaces the entity's URI throughout the interface. This is useful for usability, as users see meaningful bibliographic references like "Peroni & Shotton. OpenCitations, an infrastructure organization for open scholarship" instead of technical URIs like "https://w3id.org/oc/meta/br/062501777134".
- The `displayProperties` section defines which properties of Journal Articles should be shown in the interface and how they should be presented:
  - The RDF type property is displayed as "Type" but does not support search functionality.
  - The title property (`dcterms:title`) is configured as a textarea input with search capabilities. The `supportsSearch: true` setting enables disambiguation during data entry: when adding a title, the system will search for existing entities with similar titles to prevent duplication. The `minCharsForSearch: 2` setting requires at least two characters to be typed before search begins.
  - The identifier property uses a SPARQL query to format identifiers by combining their scheme (e.g., "doi") and value (e.g., "10.1000/123456"). Its `searchTarget: "parent"` parameter is needed because identifiers are separate entities in the data model (of type `datacite:Identifier`), connected to entities of type `fabio:JournalArticle` via the `datacite:hasIdentifier` relation. This setting ensures that when searching for an identifier, the system returns the parent entity of type `fabio:JournalArticle` that have that identifier, not the identifier entities themselves.
  - The `partOf` property uses a SPARQL query with the `frbr:partOf+` path (where '+' indicates one or more relationship steps) to find the issue that contains the article.

It is important to note that while HERITRACE can be customized through SHACL and YAML configurations, these customizations are optional rather than required for the system to function. HERITRACE is designed to work seamlessly with existing RDF data stored in a triple store without requiring extensive configuration. When connecting HERITRACE to an existing RDF dataset, the system automatically handles the integration of data and applies its provenance management and change-tracking functionalities to all subsequent modifications. Importing existing RDF graphs is as simple as connecting HERITRACE to the triple store where the data resides: no additional data transformation or special import procedures are needed. The system will automatically discover and display the entities in the triple store based on their RDF types. While SHACL descriptions are not mandatory, they enhance the user experience by providing validation and by tailoring the editing interface to the specific data model of the collection. Without SHACL descriptions, HERITRACE will still function but may not offer the same level of data validation or customized editing forms. This flexibility makes HERITRACE adaptable to various deployment scenarios, from quick implementations with existing data to fully customized installations with comprehensive validation rules.

The SHACL and YAML configuration files used for the ParaText project are available on Zenodo (Massari 2025).

## 3.2 Provenance management and change-tracking

In the context of HERITRACE, provenance refers to the metadata associated with modifications to a dataset, documenting metadata such as when the modification occurred, who was responsible for it, what specific changes were made, and the primary source of the data involved (Gil et al. 2010). Provenance management ensures a transparent and auditable history of curatorial actions. HERITRACE employs the OpenCitations Data Model (OCDM) (Daquino et al. 2020) to implement its provenance management system, extending the PROV Ontology (PROV-O) (Lebo, Sahoo, and McGuinness 2013) with additional mechanisms to record changes in the data.

Provenance in the OCDM revolves around the concept of snapshots, each representing a particular state of an entity, such as a book or an artifact, at a specific moment in time. Snapshots are stored as instances of `prov:Entity` and are linked to prior states of the entity through the `prov:wasDerivedFrom` property. This linkage allows the reconstruction of an entity's history, providing a comprehensive view of how it evolved. HERITRACE records a set of core metadata for each snapshot to ensure complete traceability. This metadata includes the timestamp of creation (`prov:generatedAtTime`), which identifies the precise moment a snapshot was created, and, when relevant, the timestamp of invalidation (`prov:invalidatedAtTime`), which marks when a snapshot was superseded by subsequent modifications. Additionally, the metadata includes the responsible agent (`prov:wasAttributedTo`), detailing who performed the modification—whether an individual, an organization, or an automated process—and the primary data source (`prov:hasPrimarySource`), which links the modification to the external or internal source that informed the change. Together, these elements form a robust provenance framework, ensuring that all curatorial actions are fully documented, and that the dataset retains a high degree of integrity and accountability.

While provenance management focuses on contextual metadata, change-tracking addresses the content of modifications. The OCDM employs a document-based approach to change tracking applied to data, where only the differences, or deltas, between successive snapshots of an entity are stored. This method avoids the inefficiency of retaining complete backups of each version of the entity. Instead, HERITRACE records SPARQL update queries to track entity changes. Specifically, additions to an entity are captured as `INSERT DATA` queries, while removals are stored as `DELETE DATA` queries. These deltas are linked to the entity through the `oco:hasUpdateQuery` property, introduced by the OCDM (Peroni, Shotton, and Vitali 2016). This approach reduces the storage requirements significantly and simplifies the process of reconstructing previous versions of the entities. To return to the previous state, HERITRACE reverses the recorded SPARQL update queries. For example, if a triple was added to an entity, the system removes it to revert to the earlier state, and if a triple was removed, it reinstates it. This process is facilitated by a dedicated tool called the Time Agnostic Library (Massari 2024b), which automates the application and inversion of update queries, ensuring an efficient and precise restoration process.

The integration of provenance management and change tracking in HERITRACE is exemplified in Figure 5, which illustrates the lifecycle of an entity through successive modifications. Initially, the creation of an entity is captured as a snapshot, with associated provenance metadata detailing the timestamp, the responsible agent, and the primary source. As modifications occur, HERITRACE generates deltas that document the specific triples added or removed. For instance, if a new author is added to a bibliographic record, the corresponding delta is stored as an `INSERT DATA` query, while the removal of outdated information is captured as a `DELETE DATA` query.

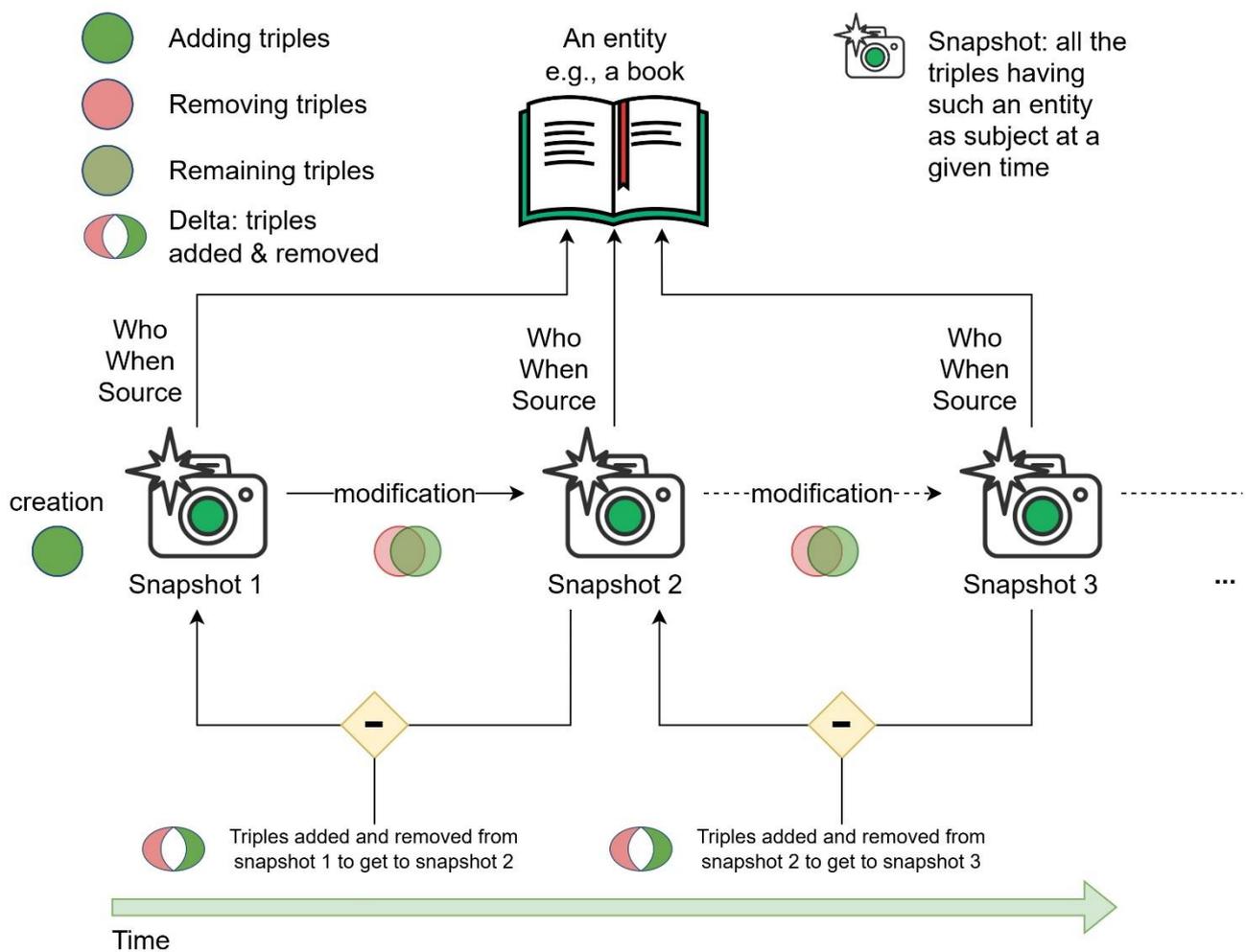

*Figure 5 OCDM change-tracking mechanism, showing the evolution of an entity through successive snapshots.*

For those interested in exploring HERITRACE further, the system along with its documentation are available on GitHub and Software Heritage (Massari 2024a).

# 4   Discussion and future directions

HERITRACE addresses key limitations identified in existing semantic data management systems by integrating usability, provenance, and change-tracking into a unified framework. While Section 2.2 provided a detailed comparative analysis of existing solutions, this section focuses on HERITRACE's approach to solving these challenges and on its future development roadmap.

By employing the OpenCitations Data Model (OCDM), HERITRACE ensures detailed provenance, recording when and by whom modifications were made, and linking changes to their primary sources. Its delta-based change-tracking approach, which utilizes SPARQL update queries, provides efficient storage and facilitates the reconstruction of previous dataset states without the need for full backups. Unlike systems that store change history in relational databases, HERITRACE maintains all data in RDF format, supporting quadruples and named graphs.

HERITRACE simplifies customization for GLAM institutions through its use of standard technologies rather than proprietary solutions. While some existing platforms require learning custom templating languages, HERITRACE employs SHACL, a W3C standard widely recognized for its role in defining and validating RDF data structures. Additionally, HERITRACE simplifies graphical customization through YAML configuration files, allowing administrators to adjust interface elements by modifying predefined parameters. This approach ensures that HERITRACE can be tailored to meet the diverse needs of GLAM institutions while remaining accessible to configure. Furthermore, HERITRACE functions effectively out of the box, providing an immediate solution for institutions without requiring extensive initial configuration.

To illustrate the practical advantages of HERITRACE's integrated approach, consider a common scenario in cultural heritage data management: correcting and tracking the provenance of bibliographic information. For example, when a researcher discovers that an author's name is incorrectly spelled in a catalog record, addressing this issue requires different workflows across platforms. In OmekaS, the curator must navigate to the item's edit page, correct the name, and save the change, but the system retains no record of who made this correction or why, nor can previous versions be easily retrieved if needed. In Semantic MediaWiki, though the correction process is straightforward, tracking the change requires activating and configuring the Semantic Watchlist extension, which stores change data in a separate database format disconnected from the semantic knowledge graph. In ResearchSpace, while the system would record who made the change and associate it with a source, there is no built-in mechanism to view or restore previous versions of the record, making temporal management of data difficult. CLEF provides robust provenance management through named graphs and integrates with GitHub for change tracking, but its web interface doesn't offer direct restoration of previous versions, requiring users to navigate between different systems. With HERITRACE, the entire process is unified: the curator makes the correction through the same intuitive interface, the system automatically documents who made the change and when, and the Time Machine feature allows any user to view the complete modification history or restore previous versions with a single click; all while maintaining the semantic integrity of the data throughout the entire lifecycle. This integration eliminates the fragmentation of workflows and the technical complexity that characterizes other solutions, significantly reducing both the learning curve and the time required for data curation tasks.

Nevertheless, HERITRACE focus on RDF is a limitation. Cultural heritage data often exists in other formats, especially tabular data (e.g., CSV files) and relational databases. This diversity creates challenges for institutions aiming to integrate legacy systems or non-RDF datasets into a unified semantic framework. To address this gap, a key future direction for HERITRACE is the integration of the RDF Mapping Language (RML) (Dimou et al. 2014). RML is a powerful extension of the W3C-standard R2RML (Relational Database to RDF Mapping Language) and enables the transformation of heterogeneous data formats into RDF. By supporting mappings from tabular data, JSON, XML, and relational databases to RDF, RML would significantly enhance the system's flexibility. Its declarative approach allows users to define mapping rules in a straightforward and reusable manner, making it an ideal solution for bridging diverse data sources. For GLAM institutions, this integration would mean that even data not natively represented in RDF could be seamlessly incorporated into HERITRACE, further increasing the system's utility.

In addition to technical enhancements, usability testing will play a pivotal role in HERITRACE's future development. We will employ task-based testing with 15-20 GLAM professionals (representing both technical and domain experts) using concurrent think-aloud protocols to capture cognitive processes during interaction with key system functions. Quantitative measurements will include task completion rates, time-on-task analysis, error frequencies, and System Usability Scale (SUS) scoring (Bangor, Kortum, and Miller 2008), while qualitative data will be gathered through semi-structured post-task interviews. This mixed-methods approach will generate both statistical performance indicators and rich contextual insights to inform iterative refinements to HERITRACE's interface architecture and interaction patterns.

Finally, a broader objective for HERITRACE is to expand its adoption and collaboration within the GLAM community. Current deployments, such as the ParaText project at the University of Bologna, demonstrate the system's potential to address real-world curatorial challenges. Moreover, HERITRACE is planned for adoption by OpenCitations (Peroni and Shotton 2020), an open infrastructure dedicated to the free dissemination of bibliographic and citation data. Both authors of this paper are involved in the OpenCitations project, which facilitates the alignment of the system with the organization's specific requirements. The adoption of HERITRACE will focus on managing metadata and citation datasets. These applications are already providing an opportunity to test HERITRACE in real-world contexts characterized by large-scale, dynamic datasets.

## 5  Conclusion

In summary, HERITRACE presents itself as a practical solution in the field of semantic data management, with a particular focus on the needs of the GLAM sector. The system provides a user-friendly interface that caters to both nontechnical and technical users, alongside features such as provenance management, change tracking, and the ability to customize according to specific needs. Its capability to integrate with existing datasets enhances its practicality. Through these efforts, HERITRACE seeks to support efficient and transparent data curation processes, enabling institutions to manage their digital collections with accuracy and flexibility.

# Acknowledgements

This work has been partially funded by Project PE 0000020 CHANGES - CUP B53C22003780006, NRP Mission 4 Component 2 Investment 1.3, Funded by the European Union - NextGenerationEU.

We would also like to thank Francesca Filograsso and her supervisor Camillo Neri from the Department of Classical Philology and Italian Studies at the University of Bologna for their work on ParaText, the first use case of HERITRACE, and for conducting guerrilla testing, raising requirements, and helping to resolve numerous bugs.

# References

ADLab - Laboratorio Analogico Digitale and /DH.arc - Digital Humanities Advanced Research Centre. 2022. "FICLIT Digital Library." Bologna, Italy: University of Bologna.

Alexiev, Vladimir. 2018. "Museum Linked Open Data: Ontologies, Datasets, Projects." *Digital Presentation and Preservation of Cultural and Scientific Heritage* 8 (September):19–50. https://doi.org/10.55630/dipp.2018.8.1.

Bangor, Aaron, Philip T. Kortum, and James T. Miller. 2008. "An Empirical Evaluation of the System Usability Scale." *International Journal of Human-Computer Interaction* 24 (6): 574–94. https://doi.org/10.1080/10447310802205776.

Beek, Wouter, Rinke Hoekstra, Fernie Maas, Albert Meroño-Peñuela, and Inger Leemans. 2014. "Linking the STCN and Performing Big Data Queries in the Humanities." In *Digital Humanities Benelux Conference 2014*.

Ben-Kiki, Oren, Clark Evans, and Brian Ingerson. 2009. "Yaml Ain't Markup Language (Yaml$^{TM}$) Version 1.1." *Working Draft 2008* 5:11.

Berthereau, Daniel and Corporation for Digital Scholarship. 2015. "CSV Import." https://omeka.org/s/modules/CSVImport/.

Candela, Gustavo. 2023. "Towards a Semantic Approach in GLAM Labs: The Case of the Data Foundry at the National Library of Scotland." *Journal of Information Science*.

Candela, Gustavo, Pilar Escobar, Rafael C. Carrasco, and Manuel Marco-Such. 2018. "Migration of a Library Catalogue into RDA Linked Open Data." Edited by Christoph Schlieder. *Semantic Web* 9 (4): 481–91. https://doi.org/10.3233/SW-170274.

Candela, Gustavo, María Dolores Sáez, MPilar Escobar Esteban, and Manuel Marco-Such. 2022. "Reusing Digital Collections from GLAM Institutions." *Journal of Information Science* 48 (2): 251–67. https://doi.org/10.1177/0165551520950246.

Carriero, Valentina Anita, Aldo Gangemi, Maria Letizia Mancinelli, Ludovica Marinucci, Andrea Giovanni Nuzzolese, Valentina Presutti, and Chiara Veninata. 2019. "ArCo: The Italian Cultural Heritage Knowledge Graph." In *The Semantic Web – ISWC 2019*, edited by Chiara Ghidini, Olaf Hartig, Maria Maleshkova, Vojtěch Svátek, Isabel Cruz, Aidan Hogan, Jie Song, Maxime Lefrançois, and Fabien Gandon, 11779:36–52. Lecture Notes in Computer Science. Cham: Springer International Publishing. https://doi.org/10.1007/978-3-030-30796-7_3.

Corine, Deliot, Wilson Neil, Costabello Luca, and Vandenbussche@ Pierre-Yves. 2017. "The British National Bibliography: Who Uses Our Linked Data?" In *Proceedings of the International Conference on Dublin Core and Metadata Applications*. Dublin Core Metadata Initiative. https://doi.org/10.23106/DCMI.952137546.

Daquino, Marilena, Silvio Peroni, David Shotton, Giovanni Colavizza, Behnam Ghavimi, Anne Lauscher, Philipp Mayr, Matteo Romanello, and Philipp Zumstein. 2020. "The

OpenCitations Data Model." In *Lecture Notes in Computer Science*, 447–63. Cham: Springer International Publishing. https://doi.org/10.1007/978-3-030-62466-8_28.

Daquino, Marilena, Mari Wigham, Enrico Daga, Lucia Giagnolini, and Francesca Tomasi. 2022. "CLEF. A Linked Open Data Native System for Crowdsourcing." https://doi.org/10.48550/ARXIV.2206.08259.

Dimou, Anastasia, Miel Vander Sande, Pieter Colpaert, Ruben Verborgh, Erik Mannens, and Rik Van de Walle. 2014. "RML: A Generic Language for Integrated RDF Mappings of Heterogeneous Data." In *Proceedings of the 7th Workshop on Linked Data on the Web*, edited by Christian Bizer, Tom Heath, Sören Auer, and Tim Berners-Lee. Vol. 1184. CEUR Workshop Proceedings. http://ceur-ws.org/Vol-1184/ldow2014_paper_01.pdf.

Gil, Yolanda, James Cheney, Paul Groth, Olaf Hartig, Simon Miles, Luc Moreau, and Paulo Pinheiro da Silva. 2010. "Provenance XG Final Report." *W3C*. http://www.w3.org/2005/Incubator/prov/XGR-prov-20101214/.

Haak, Laurel L., Martin Fenner, Laura Paglione, Ed Pentz, and Howard Ratner. 2012. "ORCID: A System to Uniquely Identify Researchers." *Learned Publishing* 25 (4): 259–64. https://doi.org/10.1087/20120404.

Hannemann, Jan, and Jürgen Kett. 2010. "Linked Data for Libraries." In *Proc of the World Library and Information Congress of the Int'l Federation of Library Associations and Institutions (IFLA)*.

Heibi, Ivan, Silvio Peroni, and David Shotton. 2019. "Enabling Text Search on SPARQL Endpoints through OSCAR." Edited by Alejandra Gonzalez-Beltran, Alejandra Gonzalez-Beltran, Francesco Osborne, Silvio Peroni, and Sahar Vahdati. *Data Science* 2 (1–2): 205–27. https://doi.org/10.3233/DS-190016.

Hyvönen, Eero. 2020. "Linked Open Data Infrastructure for Digital Humanities in Finland." *Digital Humanities in the Nordic and Baltic Countries Publications* 3 (1): 254–59. https://doi.org/10.5617/dhnbpub.11195.

Isaac, Antoine, and Bernhard Haslhofer. 2013. "Europeana Linked Open Data – Data.Europeana.Eu." *Semantic Web* 4 (3): 291–97. https://doi.org/10.3233/SW-120092.

Kremer, Christine. 2021. "Bibliothèque nationale du Luxembourg : multiple et ouverte à tous." *Arabesques*, no. 102 (July), 26–27. https://doi.org/10.35562/arabesques.2657.

Krötzsch, Markus, Denny Vrandečić, and Max Völkel. 2006. "Semantic MediaWiki." In *The Semantic Web - ISWC 2006*, edited by Isabel Cruz, Stefan Decker, Dean Allemang, Chris Preist, Daniel Schwabe, Peter Mika, Mike Uschold, and Lora M. Aroyo, 4273:935–42. Lecture Notes in Computer Science. Berlin, Heidelberg: Springer Berlin Heidelberg. https://doi.org/10.1007/11926078_68.

Lampa, Samuel, Egon Willighagen, Pekka Kohonen, Ali King, Denny Vrandečić, Roland Grafström, and Ola Spjuth. 2017. "RDFIO: Extending Semantic MediaWiki for Interoperable Biomedical Data Management." *Journal of Biomedical Semantics* 8 (1): 35. https://doi.org/10.1186/s13326-017-0136-y.

Laurence, Corinne M. 2013. "Linked Data and the Library of Congress." *Library Philosophy and Practice*, 2–24.

Lebo, Timothy, Satya Sahoo, and Deborah McGuinness. 2013. "PROV-O: The PROV Ontology." W3C. PROV-O. 2013. http://www.w3.org/TR/2013/REC-prov-o-20130430/.

Massari, Arcangelo. 2024a. "HERITRACE (Heritage Enhanced Repository Interface for Tracing, Research, Archival Curation, and Engagement)." https://archive.softwareheritage.org/swh:1:snp:4d1d83b7043649a21900fcbf6465f0879672228e;origin=https://github.com/opencitations/heritrace.

———. 2024b. "Opencitations/Time-Agnostic-Library: 3.6.13." *Software Heritage Archive*. https://archive.softwareheritage.org/swh:1:snp:9d4ed12adc4b2bdfe96f9629cb99b1ab3c559261;origin=https://github.com/opencitations/time-agnostic-library.

Massari, Arcangelo, and Silvio Peroni. 2024. "HERITRACE: A User-Friendly Semantic Data Editor with Change Tracking and Provenance Management for Cultural Heritage Institutions." In *Quaderni Di Umanistica Digitale*. Catania: AMSActa. https://doi.org/10.48550/arXiv.2402.00477.

Oldman, Dominic, and Diana Tanase. 2018. "Reshaping the Knowledge Graph by Connecting Researchers, Data and Practices in ResearchSpace." In *The Semantic Web – ISWC 2018*, edited by Denny Vrandečić, Kalina Bontcheva, Mari Carmen Suárez-Figueroa, Valentina Presutti, Irene Celino, Marta Sabou, Lucie-Aimée Kaffee, and Elena Simperl, 11137:325–40. Lecture Notes in Computer Science. Cham: Springer International Publishing. https://doi.org/10.1007/978-3-030-00668-6_20.

Pan, Jeff Z. 2009. "Resource Description Framework." In *Handbook on Ontologies*, 71–90. Springer.

Pareti, Paolo, and George Konstantinidis. 2022. "A Review of SHACL: From Data Validation to Schema Reasoning for RDF Graphs." In *Reasoning Web. Declarative Artificial Intelligence*, edited by Mantas Šimkus and Ivan Varzinczak, 13100:115–44. Lecture Notes in Computer Science. Cham: Springer International Publishing. https://doi.org/10.1007/978-3-030-95481-9_6.

Peroni, Silvio, and David Shotton. 2020. "OpenCitations, an Infrastructure Organization for Open Scholarship." *Quantitative Science Studies* 1 (1): 428–44. https://doi.org/10.1162/qss_a_00023.

Peroni, Silvio, David Shotton, and Fabio Vitali. 2016. "A Document-Inspired Way for Tracking Changes of RDF Data." In *Detection, Representation and Management of Concept Drift in Linked Open Data*, edited by L. Hollink, S. Darányi, A.M. Peñuela, and E. Kontopoulos, 26–33. Bologna: CEUR Workshop Proceedings. http://ceur-ws.org/Vol-1799/Drift-a-LOD2016_paper_4.pdf.

Petz, Georg. 2023. "Linked Open Data. Zukunftsweisende Strategien." *Bibliothek Forschung Und Praxis* 47 (2): 213–22. https://doi.org/10.1515/bfp-2023-0006.

Rachinger, Johanna. 2008. "The Austrian National Library: A New Orientation in the Shadows of a Long History." *Alexandria: The Journal of National and International Library and Information Issues* 20 (1): 151–60. https://doi.org/10.1177/095574900802000105.

Ranjgar, Babak, Abolghasem Sadeghi-Niaraki, Maryam Shakeri, Fatema Rahimi, and Soo-Mi Choi. 2024. "Cultural Heritage Information Retrieval: Past, Present, and Future Trends." *IEEE Access* 12:42992–26. https://doi.org/10.1109/ACCESS.2024.3374769.

Salarelli, Alberto. 2016. "Gestire piccole collezioni digitali con Omeka: l'esperienza di MoRE (A Museum of REfused and unrealised art projects)." *Bibliothecae.it* Vol 5 (November):177-200 Paginazione. https://doi.org/10.6092/ISSN.2283-9364/6393.

Silveira, Lúcia da, Fabiano Couto Corrêa da Silva, Sara Caselani Zilio, and Larissa Silva Cordeiro. 2020. "Convergência de Práticas Linked Open Data Na Bibliothèque Nationale de France (BNF DATA)." *Revista ACB: Biblioteconomia Em Santa Catarina* 25 (1): 21–40.

WikiTeq. 2022. "Semantic Watchlist." https://www.mediawiki.org/wiki/Extension:Semantic_Watchlist.

Wilkinson, Mark D., Michel Dumontier, IJsbrand Jan Aalbersberg, Gabrielle Appleton, Myles Axton, Arie Baak, Niklas Blomberg, et al. 2016. "The FAIR Guiding Principles for Scientific Data Management and Stewardship." *Scientific Data* 3 (1): 160018. https://doi.org/10.1038/sdata.2016.18.

Wulff, Enrique. 2024. "Digital Humanities: The Case Study of the National Library in Spain." In *Advances in Information Quality and Management*, edited by Mehdi Khosrow-Pour, D.B.A., 1–19. IGI Global. https://doi.org/10.4018/978-1-6684-7366-5.ch052.